\documentclass[a4paper,12pt]{iopart}
\usepackage{times}

\newcommand{\eq}{\begin{equation}}
\newcommand{\eqa}{\begin{eqnarray}}
\newcommand{\en}{\end{equation}}
\newcommand{\ena}{\end{eqnarray}}
\newcommand{\enn}{\nonumber \end{equation}}

\def\sk{\vskip .4cm}

\def\al{\alpha}

\def\be{\beta}

\def\we{\wedge}
\def\th{\theta}
\def\de{\delta}

\def\part{\partial}

\def\n2{{{N+1} \over 2}}

\def\square{{\,\lower0.9pt\vbox{\hrule \hbox{\vrule height 0.2 cm
\hskip 0.2 cm \vrule height 0.2 cm}\hrule}\,}}

\def\Q.E.D.{\rightline{$\Box$}}

\def\Rcal{{\cal R}}

\def\Gcal{{\cal G}}

\def\lb{\label}

\begin{document}
\title{Yang-Mills and Born-Infeld actions
on finite group spaces}
\author{{P Aschieri}${}^{1}$, {L Castellani}${}^{2}$ and {A P Isaev
  }${}^3$}
\address{${}^1$ Sektion Physik der
Ludwig-Maximilians-Universit\"at, Theresienstr. 37, D-80333
M\"unchen, Germany}
\address{ ${}^2$ Dipartimento di Scienze,
 Universit\`a del Piemonte Orientale, Dipartimento di
Fisica Teorica and I.N.F.N. Via P. Giuria 1, 10125 Torino,
Italy.}
\address{  ${}^3$ Bogoliubov Laboratory of Theoretical
Physics, JINR,  980 Dubna, Moscow Region, Russia}
\begin{abstract}
Discretized nonabelian gauge theories living on finite group
spaces $G$ are defined by means of a geometric action $\int Tr~ F
\we *F$. This technique is extended to obtain a discrete version of
the Born-Infeld action.{\footnote{Talk presented by P Aschieri.
We thank Chiara Pagani for useful discussions.
Work partially supported by a Marie Curie Fellowship
of the EC  programme IHP, contract number MCFI-2000-01982.}}
\end{abstract}

%%%%%%%%%%%%%%%%%%%%%%%%%%%%%%%%%%%%%%%%%%%%%%%%%%%

\section{Introduction}

The base space in a gauge theory is usually a manifold; we
consider here a finite group $G$ as base space, and the Lie group
$\Gcal=U(N)$ as fiber. One can associate to $G$ a ``manifold''
structure by constructing on $G$ a differential calculus. This can
be done following the general procedure first studied by
Woronowicz \cite{wor} in the noncommutative context of quantum
groups. The differential calculus on $G$ not only determines the
``manifold'' $G$, but
%(e.g. the dimension of the tangent space)
has been proven to be a very useful tool in the formulation  of
gauge and gravity theories with base space $G$
\cite{DMGcalculus,gravfg,majidetal,I,DM}. Following \cite{I}, we here
show that given a differential calculus on $G$, a Yang-Mills
action can be naturally constructed using just geometric objects:
differential forms, invariant metric, *-Hodge operator, Haar
measure. We similarly construct  a discretized version of
Born-Infeld theory. These actions can be generalized \cite{I} to
the case where the base space is $M^D\times G$, with $M^D$ a
continuous $D$-dimensional manifold. Then one can study gauge
theories on $M^D$ that are ``Kaluza-Klein'' reduced from those on
$M^D\times G$: one has just to reinterpret the $M^D\times G$ gauge
action as a new $M^D$ one. For example, approximating the circle
$S_1$ with the cyclic group $Z_n$ we obtain on $M^D$ a gauge
action plus extra scalar fields from a pure gauge action on
$M^D\times Z_n$. This provides an alternative and different
procedure from the usual Kaluza-Klein massive modes truncation in
$M^D\times S_1\rightarrow M^D$. In some cases one can obtain a
Higgs field and potential; a similar mechanism is present in the
Connes-Lott standard Model \cite{Connes}.

Noncommutative structures in string/brane theory have emerged in
the last years and are the object of intense research. Our study
is in this general context. The noncommutativity we discuss is
mild, in the sense that fields commute between themselves (in the
classical theory), and only the commutations between fields and
differentials, and of differentials between themselves, are
nontrivial. This noncommutativity is in a certain sense
complementary to the one of  $\star$ (Moyal)-deformed theories,
where (for constant $\th$) only functions do not commute. It is
tempting to interpret the product space $M^D\times G$ as a bundle
of $n$ (D-1)-dimensional branes evolving in time, $n$ being the
dimension of the finite group $G$.

The paper is organized as follows. In Section 2 we recall some basic
facts about the differential geometry of finite  groups,
and construct those geometric tools needed in Section 3
for the study of Yang-Mills theories.
In Section 4 an action for the  Born-Infeld theory is presented.

\section{Differential geometry on finite groups}

 Let $G$ be a finite group of order $n$ with
generic element $g$ and unit $e$. Consider $Fun(G)$, the set of
complex functions on $G$.
The left and right actions of the group $G$ on $Fun(G)$ read
$
\lb{dg2b} {\cal L}_{g} f|_{_{g'}} =f( g \, g')\,,~{\cal R}_{g\,}
f|_{_{g'}}= f( g' \, g)~\; \forall f \in Fun(G)\, .
$
The vectorfields $t_{g}$ on $G$ are defined by their action
on $Fun(G)$:
$
t_gf|_{_{g'}}=f(g'g)-f(g')=({\cal R}_{g}f-f)|_{_{g'}}
$,
with $g\not=e$. They are nonlocal and left-invariant:
${\cal L}_{g'} (t_gf)=t_g({\cal L}_{g'}f)$. We also have
${\cal R}_{g'} (t_gf)=t_{g'g{g'}^{-1}}({\cal R}_{g'}f)$.
The differential of an arbitrary function $f \in Fun(G)$
is then given by
 \eq
d f =\sum_{g \neq e} \,  (
t_g \, f) \, \theta^g \;~,  \label{partflat}
 \en
where $\th^g$ are one-forms; by definition they span
the linear space dual to that of the vectorfields $t_g$:
$\langle t_g,\th^{g'}\rangle=\delta^{g'}_g$.
{}From the Leibniz rule for the differential
$ d(fh)=(df)h+f(dh),~~\forall f,h\in Fun(G)$,
we have that functions do not commute with one-forms:
\eq
   \theta^g f = (\Rcal_g f) \theta^g~~~~~~~~~(g \not= e)~.
    \label{fthetacomm}
\en
Similarly, left and right invariance of the differential
${\cal L}_g \, df=d({\cal L}_gf)~,~{\cal R}_g \, df=d({\cal R}_gf)$
implies that $\th^g$ are left-invariant one-forms
${\cal L}_{g'} (\th^g)=\th^g$
and that
\eq
{\cal R}_{g'} (\th^g)=\th^{g'g_{\,}{g'}^{-1}}~.
\label{Rtheta}
\en
A generic one form $\rho$ can always be written in the $\th^g$ basis as
$\rho=\sum_{g\not=e}f_g\th^g$, with $f_g\in Fun(G)$. We have here
described the so-called universal differential calculus on
$G$. Smaller calculi can be obtained setting $\th^g=0$ for some
$g\in G$. Because of (\ref{Rtheta}), the new differential is
still left and right invariant iff, given $\th^g=0$, also
$\th^{g'g{g'}^{-1}}=0$.
In other words, (bicovariant) differential calculi are
in 1-1 correspondence with unions of coniugacy classes of $G$.

The algebra $Fun(G)$ has a natural *-conjugation:
$f^*(g)\equiv\overline{f(g)}$, where $\overline{{~}^{^{{}^{~}}}}$
is complex conjugation. This *-conjugation can be extended to the
differential algebra, the reality condition $(df)^*=d(f^*)$ then
implying
 $(\th^g)^*=-\th^{g^{-1}}$. If we set $\th^g=0$ then  the
one-forms in both the $\{\th^{g}\}$ and the $\{\th^{g^{-1}}\}$
conjugacy classes should be set to zero.

An exterior product, compatible with the left and right
actions of $G$, can be defined as (here $\otimes$ by definition satisfies
$\rho f\otimes \rho'=\rho \otimes f\rho'$ with $\rho,\rho'$
generic one-forms)
\eq
 \theta^g \wedge \theta^{g'}
= \theta^g \otimes \theta^{g'}
-  \theta^{g g' g^{-1}} \otimes \theta^g = \theta^g
\otimes \theta^{g'} - ({\cal R}_g \theta^{g'}) \otimes \theta^g
\; , \;\;\; (g,g' \neq e) \label{extheta}~.~
\en

  The metric $<\;,\;>$  maps couples of 1-forms
  $\rho,\sigma$ into $Fun(G)$, and is required $<\;,\;>$
to satisfy the properties
$<f\rho,\sigma h>=f<\rho,\sigma>h~,\,~<\rho f,\sigma>
  =<\rho, f\sigma>\,,$ and to be symmetric on left-invariant one-forms.
 Up to a normalization the above conditions imply
 \eq
 g^{rs} \equiv <\theta^r,\theta^s> \equiv \de^r_{s^{-1}} \label{metric}~.
 \en
 We can generalize $<\!\!~,\!\!~>$  to tensor products
 of left-invariant one-forms as follows:
\eq
 < \theta^{i_1} \otimes \cdots \otimes \theta^{i_k},
  \theta^{j_1} \otimes \cdots \otimes \theta^{j_k}> \equiv
< \theta^{i_1}, \theta^{j'_1}> < \theta^{i_2}, \theta^{j'_k}>\ldots
 < \theta^{i_k}, \theta^{j_k}>  \label{pairingtensors}~
\en
where $j'_p=(i_{p+1}i_{p+2}...i_k) j_p (i_{p+1}i_{p+2}...i_k)^{-1}$,
i.e. $\th^{j'_p}=R_{i_{p+1}i_{p+2}...i_k}\th^{j_p}$.
  The pairing (\ref{pairingtensors}) is extended to all tensor
  products by $ <f\rho,\sigma h>=f<\rho,\sigma>h$ where now
$\rho$ and $\sigma$ are generic tensor products of same order. Then, using
  (\ref{fthetacomm})
and
  (\ref{pairingtensors}),  one can prove
  that $<\rho f,\sigma> =<\rho, f\sigma>$ for any function $f$.
 Moreover  $<\;,\;>$ is left and right invariant.

{}Finally, if there exists a form $vol$ of maximal degree, then it can be
chosen left-invariant, right-invariant and real.
The Hodge dual is then defined by $\rho\wedge *\sigma
=<\rho,\sigma>vol$. It is left linear; if $vol$ is central it is also
right linear: $*(f\rho\,h)=f(*\rho)h$.

%++++++++++++++++++++++++++++++
%%%%%%%%%%%%%%%%%%%%%%%%%%%%%%%%%%%%%%%%%%%%%%%%%%%%%%%%%%%%

\section{Gauge theories on finite groups spaces}

%%%%%%%%%%%%%%%%%%%%%%%%%%%%%%%%%%%%%%%%%%%%%%%%%%%%%%%%%%%%
The gauge field of a Yang-Mills
theory on a finite group $G$ is a matrix-valued one-form
$A(g)=A_h(g) \theta^h$. The components $A_h$ are  matrices whose entries
are functions on $G$, $A_h=(A_h)^\al_{\:\be}$, $\al ,\be=1,...N$.
As in the usual case, $\Gcal$ gauge transformations are given by
  \eq
  A'=-(dT)T^{-1}+TAT^{-1} \label{gaugeA}~~,
  \en
  where $T(g)=T(g)^{\al}_{~\be}$ is an
$N\times N$ representation of a $\Gcal$
 group element; its matrix entries belong to $Fun(G)$.
The 2-form field strength $F$ is given by the
familiar expression
\[F=dA+A\we A~,\]
 and satisfies the Bianchi identity:
$
 d \, F + A \we F - F \we A =  0\,.
$
Note that $A \we A \not= 0$ even if the gauge group $\Gcal$ is
 abelian. Thus $U(1)$ gauge theory on a finite group space
 looks like a nonabelian theory,
 a situation occurring also in gauge
 theories with $\star$-product noncommutativity.
Under the gauge  transformations (\ref{gaugeA}), $F$ varies homogeneously:
$
 F'=TFT^{-1}\,.
$
We also have
\eq
 F= U \we
  U \label{FasU2}~,
\en
where
$
 U_h = 1 + A_h \, , \;
U = \sum_{h \neq e} U_h  \theta^h = \sum_{h \neq e}  \theta^h
+ A \label{defU}\,.
$
{}From  (\ref{gaugeA}) we see that also $U$ transforms covariantly:
$U' = T \, U \, T^{-1} \, .$
 Defining the components $F_{h,g}$ as:
 \eq
  F\equiv F_{h,k}~ \theta^h \otimes \theta^k \label{Fcomp}
  \en
  eq. (\ref{FasU2}) yields:
  \eq
  F_{h,k} = U_h \, ({\cal R}_h \, U_k) -
  U_k ({\cal R}_k \, U_{k^{-1}hk}) \; . \label{FasUcomp}
  \en
\sk
We now have all the geometric tools needed to construct a
Yang-Mills action. The Yang-Mills action is the geometrical
action quadratic in $F$ given by
\eq
 A_{YM}=-\!\int Tr (F \we *F)=-\!\int Tr <F,F> vol
=-\!\sum_{s\in G} Tr <F,F> ~
\label{YMact} \en
the right-hand side being just the Haar measure of the function $Tr <F,F>$.
Recalling the properties of the pairing $<\;,\;>$, the proof
 of gauge invariance of $ Tr <F,F>$ is immediate:
 $Tr <F',F'>=Tr<TFT^{-1},TFT^{-1}>=Tr T<F,F>T^{-1}=Tr<F,F>~.$
\sk
The metric (\ref{metric}) is an euclidean metric (as is
seen using a real basis of one-forms) and as usual we require
(\ref{YMact}) to be real and positive definite. This restricts the
gauge group $\Gcal$ and imposes reality conditions on the gauge
potential $A$. Positivity of (\ref{YMact}) requires $-(Tr
<F,F>) \geq 0$. Explicitly
$
  <F,F> =
%<F_{r,s} \theta^r \otimes \theta^s, F_{m,n}
%\theta^m \otimes \theta^n>  =
  F_{r,s} <\theta^r \otimes \theta^s,
\theta^m \otimes \theta^n> \Rcal_{n^{-1}m^{-1}}F_{m,n} =  F_{r,s}
\Rcal_{rs} F_{s^{-1}r^{-1}s,s^{-1}}
$, and therefore
\eq -Tr   <F,F>= -(F_{r,s})^\al_{~\be} (\Rcal_{rs}
F_{s^{-1}r^{-1}s,s^{-1}})^\be_{~\al}~;\label{tracciaFF}
\en
we see that (\ref{tracciaFF}) is positive  if $(\Rcal_{rs}
F_{s^{-1}r^{-1}s,s^{-1}})^\be_{~\al}=-(F_{r,s}^*)^\al_{~\be}$.
This holds if (use (\ref{FasUcomp}))
\eq
 U=-U^\dagger~~~~~ \mbox{i.e}~~~~~~A^\dagger = - A ~; \label{hermiticity}
\en
here hermitian conjugation on matrix valued one forms
$A$ (or $U$) is defined as:
  \eq
  A^\dagger = (A_h \theta^h)^\dagger =
(\theta^h)^*
  A_h^\dagger=-\theta^{h^{-1}} A^\dagger_h=-\theta^h
  A^{\dagger}_{h^{-1}}=-({\cal R}_h A^\dagger_{h^{-1}})\,\th^h ~.
  \en
{}Finally gauge trasformations must preserve antihermiticity of
$A$, and this is the case if the representation $T$ of $\Gcal$ is
unitary. We thus obtain that the action (\ref{YMact}) has maximal
gauge group $\Gcal=U(N)$. Notice that writing
$A=(A_h){}^\al_{\;\be}\th^h$, where $\al,\be=1,...N$, the reality
condition $A^\dagger=-A$ is equivalent to
$\overline{A_h(g)}{}^\be_{\;\al}=A_{h^{-1}}(hg)^\al_{\;\be}\,$ and
is thus not local (not fiberwise). It follows that $A$ has values
in $M_{N\times N}({\mbox{\bf{C}}})$ and not in the Lie algebra of $U(N)$.
Nevertheless  $A^\dagger=-A$ is a good reality condition because it cuts
by half the total number of components of $A$. This can be seen
counting the real components of the antihermitian field $A$: they are
$N^2\times n\times d$, where $n$ is the number of
points of $G$, and the ``dimension'' $d$ counts the number of independent
left-invariant one-forms.
In conclusion, when $\Gcal=U(N)$, we have a bona fide pure gauge
action, where the number of components of $A$ is consistent with
the dimension 
of the gauge group.
\sk
The action (\ref{YMact}) can be expressed in terms of the link fields
 $U_h$: substituting (\ref{Fcomp}), (\ref{FasUcomp}) into (\ref{YMact})
leads to:
$ A_{YM} = 2 \sum_G Tr [U_k U^\dagger_k U_h U^\dagger_h-U^\dagger_k U_h
 (\Rcal_h U_k)(\Rcal_k U^\dagger_{k^{-1}hk})] \label{YMactionU}
$.
We could as well start with a Yang-Mills action on $M^D\times G$
and then obtain ($\mu,\nu=1,...D$)
 \eqa
  & &A_{YM} =- \int_{M^D \times G} Tr~F \we *F
= -\int_{M^D} d^Dx \sum_{G}~ Tr~[F_{\mu\nu}
  F_{\mu\nu} \nonumber\\
& &~~~~~~~~~~~
{}~~~~~ ~~~~+ {1\over 2} D_\mu U_k (D_\mu U_k)^\dagger
    +2 U_k U^\dagger_k U_h U^\dagger_h-2 U^\dagger_k U_h
 (\Rcal_h U_k)(\Rcal_k U^\dagger_{k^{-1}hk})]
   \nonumber
\ena
 This action describes
  a Yang-Mills theory on $M^D$ minimally coupled to
  the scalar fields $U_g$, with a nontrivial quartic potential.

%%%%%%%%%%%%%%%%%%%%%%%%%%%%%%%%%%%%%%%%%%%%%%%%%%%%%%%%

\section{Born-Infeld theory on finite group spaces}

%%%%%%%%%%%%%%%%%%%%%%%%%%%%%%%%%%%%%%%%%%%%%%%%%%%%%%%%

Due to space limitations we only consider here a Born-Infeld
action on abelian finite group space $G$. In the commutative case
we have
$
 A_{BI} =  \int_{M^D} d^Dx \sqrt{ \det ( \delta_{\mu\nu} +
 F_{\mu\nu})}\,.
$
The analogue of $\de_{\mu\nu}+F_{\mu\nu}$ becomes simply
$E_{g,h} \equiv \de_{g,h^{-1}} + iF_{g,h}$
(the $i$ is because a constant $F_{g,h}$ 
is an antihermitian matrix if $g=g^{-1}$ and $h=h^{-1}$).
The matrix $E_{g,h}$ transforms under $U(N)$
gauge variations in the same way
as $F_{g,h}$: $\,E'_{g,h}=TE_{g,h}R_{gh}T^{-1}$.
We need now a gauge covariant definition of determinant for a matrix
transforming as $E_{g,h}$. We define 
%the determinant of $E_{g,h}$ as:
\eqa
  {\det}_G E_{g,h }&=& \
  \epsilon^{g_1, \dots, g_p} ~ E_{g_1,h_1} ~ ({\cal
R}_{g_1 h_1} E_{g_2,h_2} ) ~ ({\cal R}_{g_1 h_1 g_2 h_2}
 E_{g_3,h_3}) \dots\nonumber\\
& &~~~~~~~~~~~~~~~~~~~~~~~~~~~~~~~~~~~~~~~\dots ({\cal R}_{g_1 h_1 g_2 h_2 \dots g_{p-1}
h_{p-1}} E_{g_p,h_p})
 ~ \epsilon^{h_1, \dots, h_p}  \nonumber\\
 & \equiv  &\epsilon^{g_1, \dots, g_p} ~
 E_{g_1,h_1,g_2,h_2,\dots , g_p,h_p}  ~ \epsilon^{h_1, \dots,
h_p} \label{detG}~\,, \ena where $\epsilon^{h_1, \dots,h_p}$ is
the usual antisymmetric epsilon tensor (this definition can be
generalized to the case of nonabelian $G$ \cite{I}). Then one can
prove that
$
 {\det}_G E'_{g,h}= T~ {\det}_G E_{g,h} ~T^{-1}~.
$
This determinant is also real and the discrete Born-Infeld action
reads
  \[
  A^G_{BI} = \int_G Tr~\sqrt{{\det}_G ~(\de_{g,h^{-1}} + iF_{g,h})}
  ~vol =
   \sum_G Tr~ \sqrt{ {\det}_G~ (\de_{g,h^{-1}} + iF_{g,h})}~~.
   \]

\end{document}